\documentclass[]{aa}  
\pdfoutput=1
\usepackage{graphicx}
\usepackage{txfonts}
\begin{document}

   \title{Globular cluster chemistry in fast rotating dwarf stars\\ belonging to
   intermediate age open clusters}



   \author{E.~Pancino
          \inst{1}
          }

   \institute{INAF -- Osservatorio Astrofisico di Arcetri, Largo Enrico Fermi 5,
   50125 Firenze, Italy}

   \date{Received ...; accepted ...}

 
  \abstract{The peculiar chemistry observed in the multiple populations of
  Galactic globular clusters is not generally found in other systems like dwarf
  galaxies and open clusters, and at the moment no model can fully explain its
  presence. Exploring the boundaries of the multiple population phenomenon and
  the variation of its extent in the space of cluster mass, age, metallicity,
  and compactness has proven to be a fruitful line of investigation. In the
  framework of a larger project to search for multiple population in open
  clusters, based on literature and survey data, I found peculiar chemical
  abundance patterns in a sample of intermediate age open clusters with publicly
  available data. More specifically, fast rotating dwarf stars ($v \sin i
  \geq$50~km~s$^{-1}$) belonging to four clusters (Pleiades, Ursa Major, Come
  Berenices, Hyades) display a bimodality in either [Na/Fe], [O/Fe] or both,
  with the low-Na and high-O peak more populated than the high-Na and low-O
  peak. Additionally, two clusters show a Na-O anticorrelation in the fast
  rotating stars and one cluster shows a large [Mg/Fe] variation among the stars
  with high [Na/Fe], reaching the extreme Mg depletion observed in NGC~2808.
  Even considering that the sample sizes are small, these patterns call for
  attention in the light of a possible connection with the multiple population
  phenomenon of globular clusters. The specific chemistry observed in these fast
  rotating dwarf stars is thought to be produced by a complex interplay of
  different diffusion and mixing mechanisms, like rotational mixing and mass
  loss, in turn influenced by metallicity, binarity, mass, age, variability, and
  so on. However, with the sample in hand, it was not possible to identify which
  stellar parameters cause the observed Na and O bimodality and Na-O
  anticorrelation, suggesting that other stellar properties might be important
  besides stellar rotation. Stellar binarity might influence the rotational
  properties and enhance rotational mixing and mass loss of stars in a dense
  environment like that of clusters (especially globulars). In conclusion,
  rotation and binarity appear as a promising research avenue to better
  understand multiple stellar populations in globular clusters, that is
  certainly worth exploring further.}

   \keywords{Stars: abundances -- globular clusters: general -- open clusters
   and associations: general -- Stars: rotation -- binaries: general}

   \maketitle
%

\section{Introduction}
\label{sec:intro}

The long standing problem of Multiple Populations (MPs) in Globular Clusters
(GCs) is still awaiting for a solution. Briefly, GC stars -- that have not
undergone classical chemical evolution like in dwarf galaxies -- display
prominent abundance variations among light elements, that often take the form of
anti-correlations. The most widely studied anti-correlations are the C-N, Na-O,
and Mg-Al ones \citep{kraft94,gratton12}. Variations are also observed in helium,
lithium, flourine, potassium, and s-process elements
\citep{smith05,strader15,dorazi15}. Photometry reflects these abundance
variations in the form of multiple photometric sequences, that are at the moment
not fully explained \citep{sbordone11,milone17}.

The observed chemical patterns are generally ascribed to hydrogen burning
through the CNO cycle and hotter Ne-Na and Mg-Al cycles \citep{denisenkov89}.
Different scenarios were built around different possible polluting stars, like
asymptotic giant branch stars, fast rotating massive stars, massive interacting
binaries, or supermassive stars
\citep{decressin07a,decressin07b,ventura13,demink09,denissenkov14}. These are
collectively known as generational scenarios, because they postulate that an
initial stellar generation with normal halo chemistry pollutes the intracluster
gas, that in turn forms a second stellar generation, with an age difference of
$\simeq$3--200~Myr, depending on the scenario. Several other non-generational
scenarios or original ideas were put forward, but were less pursued and tested
than generational scenarios. Unfortunately, generational scenarios suffer from a
series of problems \citep[see][for more details]{renzini15,bastian17} that
are presently not solved. 

One fruitful line of investigation has been to search stellar clusters
with different properties like Open Clusters (OC) or young massive clusters to
put a boundary around the phenomenon in the space of age, mass, metallicity, and
compactness \citep{bragaglia12,krause13,cabrera16,cabrera17,martocchia18}. In
the framework of a larger project to search for MPs in OCs, I collected
literature data and found chemical patterns impressively similar to those of GC
stars in fast rotating A and F dwarfs in 100-800~Myr OCs. This paper
briefly reports and discusses the findings, calling the community attention to
physical processes that have not been thoroughly explored so far to explain MPs
in GCs.

\begin{table}
\caption{Selected clusters with basic information and literature source of ages
and spectroscopic analysis. [Fe/H] estimates are the median values from the
respective studies, and agree well with the estimates by \citet{netopil16},
within uncertainties.}
\label{tab:clusters} 
\begin{small}     
\centering                         
\begin{tabular}{l c c l}        
\hline\hline                
Cluster & [Fe/H] & Age & Data source \\  
        & (dex) & (Myr) & \\ 
\hline                       
Pleiades       &  +0.12 & 115 & \citet{gebran08b} \\
Ursa Major     & --0.08 & 500 & \citet{monier05} \\
Coma Berenices &  +0.07 & 591 & \citet{gebran08a} \\
Hyades         &  +0.08 & 625 & \citet{gebran10} \\
Praesepe       &  +0.13 & 729 & \citet{fossati07,fossati08} \\
\hline                                   
\end{tabular}
\end{small}
\end{table}

   \begin{figure*}
   \raggedleft
   \includegraphics[width=\textwidth]{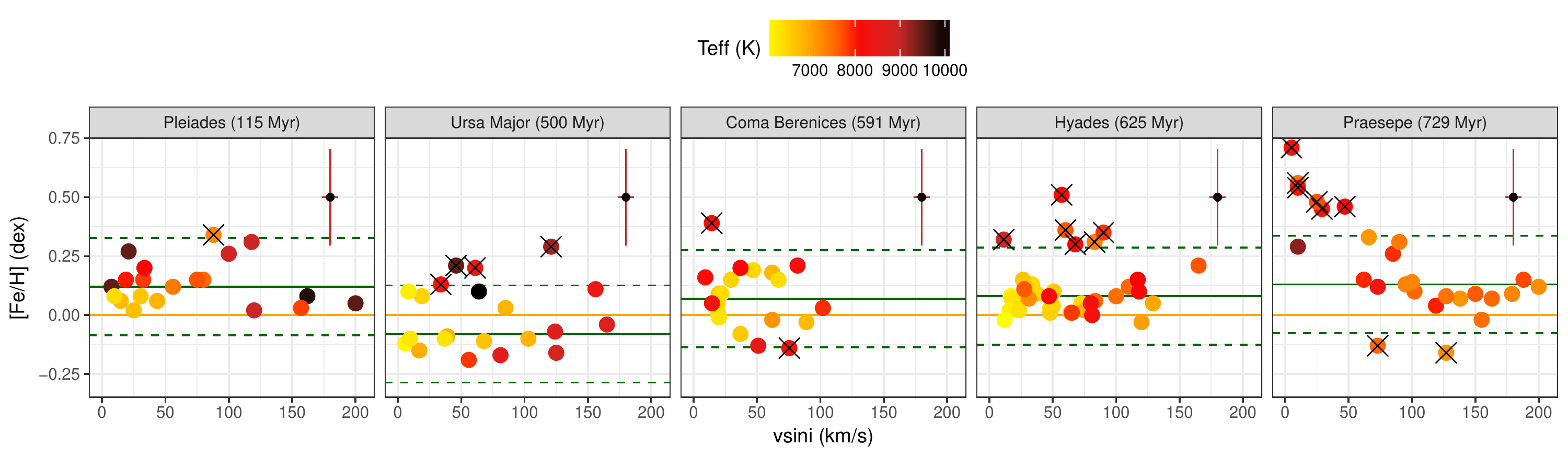}
   \includegraphics[width=\textwidth]{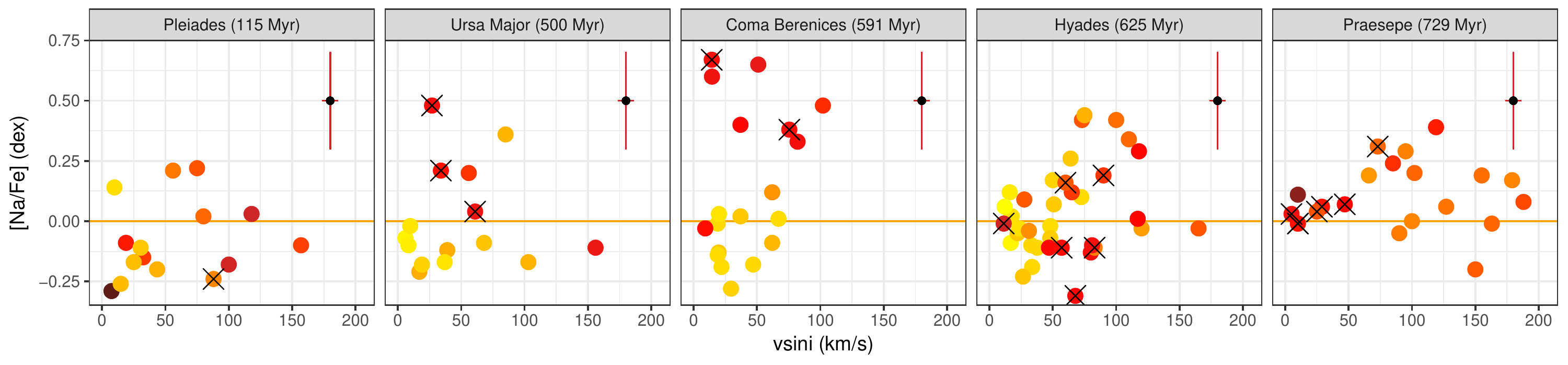}
   \includegraphics[width=\textwidth]{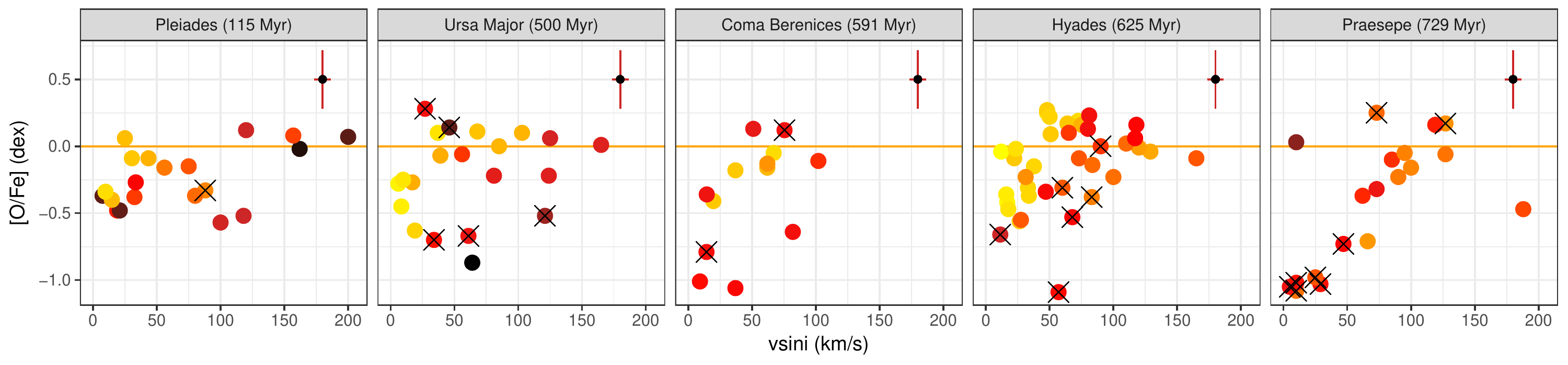}
   \includegraphics[width=\textwidth]{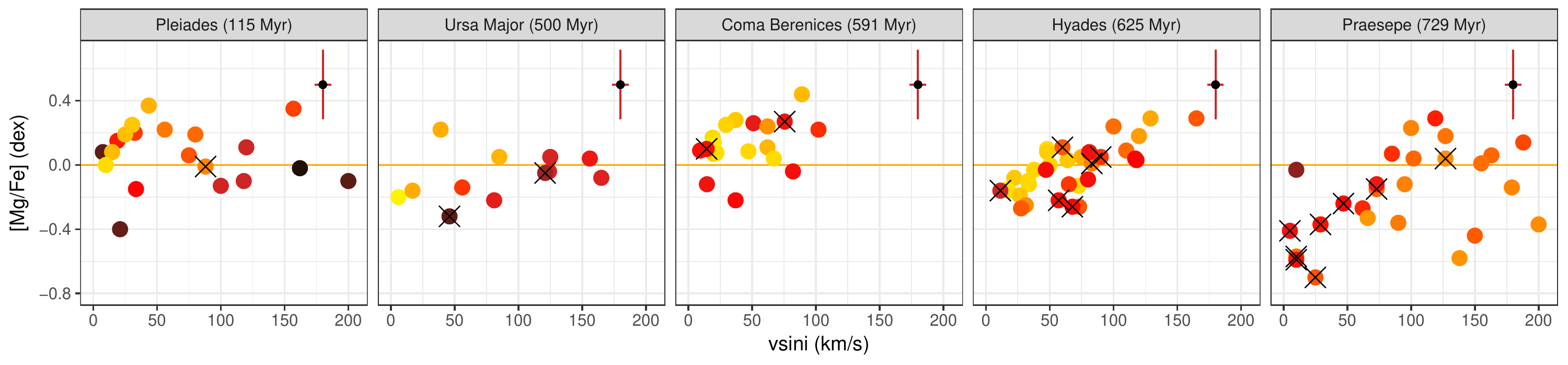}
   \caption{Behaviour of the elements of interest as a function of $v \sin
   i$. Each row displays a different element and each column a different OC,
   sorted by age, as annotated. Stars are coloured as a function of
   T$_{\rm{eff}}$. The median [Fe/H] abundance of each OC is represented as a
   solid green line in the top panels, along with its error range (dashed green
   lines). Stars with [Fe/H] outside the normal error range, suspected of being
   chemically peculiar Am and Fm stars affected by strong metallicism, are
   marked with black crosses. The solar line is plotted in orange. Median
   errorbars are plotted in dark red.} 
   \label{fig:vsini}
    \end{figure*}


\section{Data}
\label{sec:data}

Abundance measurements of A and F dwarfs in five extremely well studied clusters
of different ages were collected from the literature, as indicated in
Table~\ref{tab:clusters}. All measurements of the projected rotational velocities,
stellar parameters, and element abundances were obtained from high-quality echelle
spectra, with R$\simeq$30\,000--75\,000 and S/N$\simeq$100--600. The abundance
analysis was performed by various teams that used different methods, models,
line-lists, solar reference abundances and so on. Whenever possible, I preferred
works using similar methods, although offsets of the order of 0.1~dex between
one study and the other are always to be expected. All methods were specifically
developed for the treatment of fast rotating stars, are strictly based on spectral
synthesis, and the cited papers include the discussion of non-LTE effects.
Different works on the same cluster generally agree with each other within the
reported uncertainties and present detailed and satisfactory comparisons with
previous literature. 

In particular, the Hyades A and F dwarfs were studied by both \citet{gebran10}
and \citet{varenne99} with comparable outcomes. The \citet{gebran10} sample was
preferred because of the higher spectral S/N ratio (200--600) and because the
abundance analysis method was the same employed for the chosen Pleiades and Coma
Berenices datasets \citep{gebran08a,gebran08b}. Similarly, the
\citet{fossati07,fossati08} analysis of Praesepe agrees with past work
\citep{hui97,hui98,andrievsky98,burkhart98}, but is based on a larger sample,
with higher S/N, and more homogeneous analysis with the other selected
literature sources.

The final collected sample contains 105 well-known stars with 6$<v \sin i
<$200~km~s$^{-1}$\footnote{Note that $v \sin i$ has to be considered as a
lower limit to the actual rotational velocity.}, all from the Henry Draper
catalogue \citep{cannon18} and with extremely well studied properties in the
literature. The behaviour of the four elements of interest as a function of
$v \sin i$ is displayed in Figure~\ref{fig:vsini}\footnote{Abundance ratios
in all figures are computed using the [Fe/H] provided for each star by the
respective authors.}. Only stars considered as bona-fide members by the
respective authors were retained. The samples contain also a few stars in
binaries (where the companion does not contaminate the spectra significantly),
some $\delta$~Scuti variables, a couple of blue stragglers, and several peculiar
Am and Fm stars \citep[see][and Section~\ref{sec:other} for more
details]{alecian13}. These peculiar stars do not occupy clearly distinct
positions in the space of the elements analyzed here, with the exception of
stars with large [Fe/H] variations, as indicated in Figure~\ref{fig:vsini},
which are suspected or confirmed Am and Fm stars affected by strong metallicism.
These stars were removed because it is well known that they do not exist in GCs.
Theoretically, the effects of diffusion mechanisms are expected to be much
smaller \citep{richard02} for population~II stars, and this is confirmed
observationally \citep{korn07}. Iron variations are generally lower than
0.05~dex in the vast majority of GCs \citep{mucciarelli15}. Therefore these
peculiar stars were removed from the sample and will not be considered in the
following.

   \begin{figure*}
   \raggedleft
   \includegraphics[width=\textwidth]{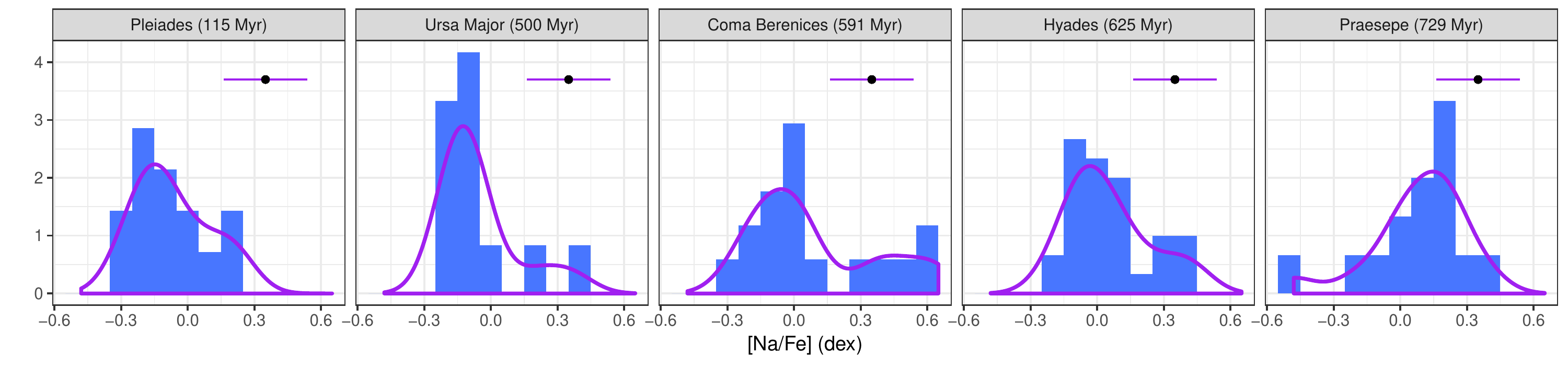}
   \includegraphics[width=\textwidth]{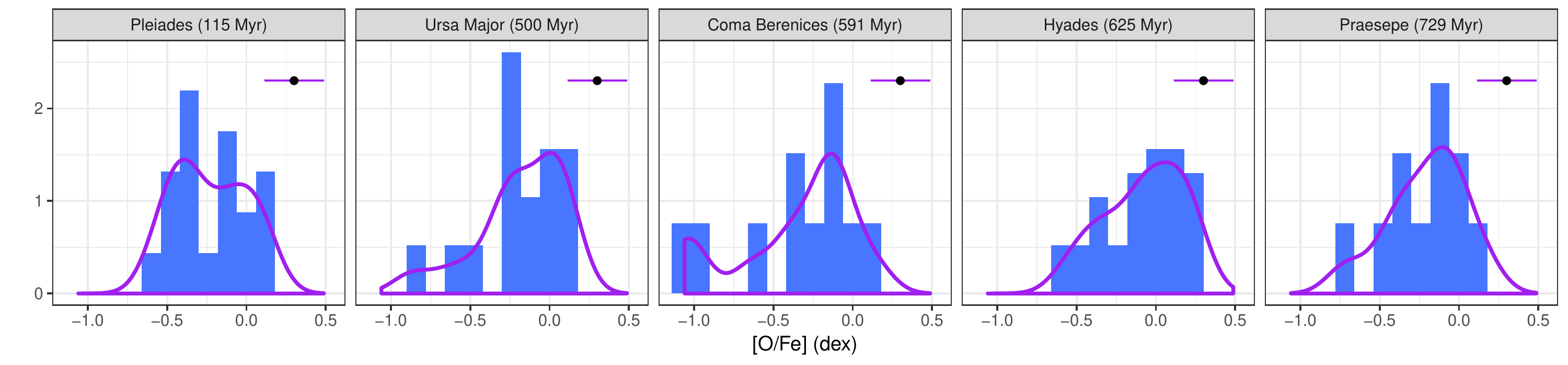}
   \caption{The [Na/Fe] distribution is represented in the top panels --
   one for each OC -- as a binned histogram (light blue bars) and a generalized
   histogram (purple lines). The smoothing kernel width was set equal to the bin
   size, i.e., 0.1~dex. The typical (median) errorbar is also
   reported in each panel. The bottom panels report the same histograms, but for
   [O/Fe].}  
   \label{fig:hist}
    \end{figure*}

The common conclusion of all the cited studies was that A stars show an
increased abundance spread in all elements compared to F stars. At these ages
and metallicities, A stars are tendentially faster rotators than F stars.
Comparisons with diffusion models for F stars \citep[generally those
by][]{turcotte98} showed that additional mixing mechanisms must be operating in
early F and in A stars, preventing the expected decrease in light elements and
increase in heavy elements. Because the disagreement with models generally
increases with $v\sin i$, rotational mixing was proposed by all considered works
as the most important of all mixing phenomena, to restore agreement with the
observations. 

Purely rotational models from the Geneva group \citep[see][and references
therein]{lagarde12} do not reproduce the observed abundance variations, and
foresee variations of $<$0.1~dex in C, N, O, Na, Mg, and Al for these dwarfs.
However, rotation does have the potential of competing with diffusion processes,
and also to help explaining the peculiar Am stars chemistry \citep{talon06}.
Works from the Montreal group further explored the capability of turbulent
mixing and of mass loss \citep[see][for example]{vick10,michaud11} as competing
mechanisms against diffusive processes and found that they both can explain
observations. At the moment, therefore, the exact mechanism that counteracts the
expected diffusion effects is not univocally identified, but there are various
equally valid possibilities that are difficult to discriminate
observationally.


\section{Results}
\label{sec:res}

\subsection{Bimodal distributions of [Na/Fe] and [O/Fe]}
\label{sec:hist}

The first surprising result is an apparent [Na/Fe] bimodality in all clusters
except Praesepe. This is visible in all figures, but is more noticeable in
Figure~\ref{fig:hist} (top panels), where the [Na/Fe] histogram is displayed for
each of the five OCs. The [Na/Fe] distribution has a spread that is
significantly larger than the typical (median) errorbar. The peak width is
roughly compatible with the typical uncertainties, while the peak separation is
generally larger than the typical uncertainties. A Gaussian mixture model
fit to the [Na/Fe] distribution, with varying number of components, clearly
provides the best BIC (Bayesian Information Criterion) with two Gaussians,
except for the Pleiades and Praesepe. Using equal or variable variance models
does not change the result, because the best fitting Gaussians have compatible
variances in all cases. Generally, the left peak (low [Na/Fe]) contains more
stars than the right peak.

A bimodality is also suggested by visually inspecting the [O/Fe]
distributions (Figure~\ref{fig:hist}, bottom panels): two peaks are generally
apparent and their width and separation are compatible with the typical
uncertainties. The relative importance of the two visually apparent peaks is
generally reversed compared to the [Na/Fe] distributions. However, the data
points are less numerous and the histograms noisier: two OCs only, the Pleiades
and Coma Berenices, have higher BIC with two-Gaussian fits than with
one-Gaussian fits. Finally, the distribution of [Mg/Fe] shows no clear
bimodality.

Unfortunately, I could not identify univocally the parameters that govern
the bimodality with the sample in hand. In Coma Berenices there is clearly a
T$_{\rm{eff}}$ difference between the two peaks (see also
Figure~\ref{fig:vsini}), but the range in $v \sin i$ is not large and both
groups contain stars of varying $v \sin i$. The clear T$_{\rm{eff}}$ difference
between the peaks that is visible in Coma Berenices is not visible in any of the
other OCs, except maybe for Ursa Major (see Figure~\ref{fig:vsini}). The
hottests stars in Coma Berenices are often classified as Am stars but in the
other OCs the Am and Fm stars are not confined to one of the two peaks. In the
Hyades, stars in binaries tend to be often -- but not always -- on the upper
sequence, but this does not happen so clearly in the other OCs.

None of the theoretical models mentioned in the preceding section foresee
such bimodality, if the stellar parameters are the same. Therefore, it is
necessary to understand what stellar property assigns stars to each of the two
peaks, if the peaks are confirmed to be distinct. Sample size is not the only
thing that needs to be improved to further investigate the matter: it will be
necessary to build samples that are as unbiased as possible with respect to the
relevant or interesting parameters, like binarity, peculiarity, variability and
pulsation, rotation, temperature, and so on. In particular, the range of
T$_{\rm{eff}}$ covered in Praesepe is smaller than in Coma Berenices, while the
range in $v \sin i$ is smaller in Coma Berenices than Praesepe. While some of
these differences might be intrinsic, and thus unavoidable, we need to have more
understanding of the data properties, and more controlled data samples, before
deriving any further conclusions.

\subsection{Na-O anticorrelation}

\begin{figure*}
   \raggedleft
   \includegraphics[width=\textwidth]{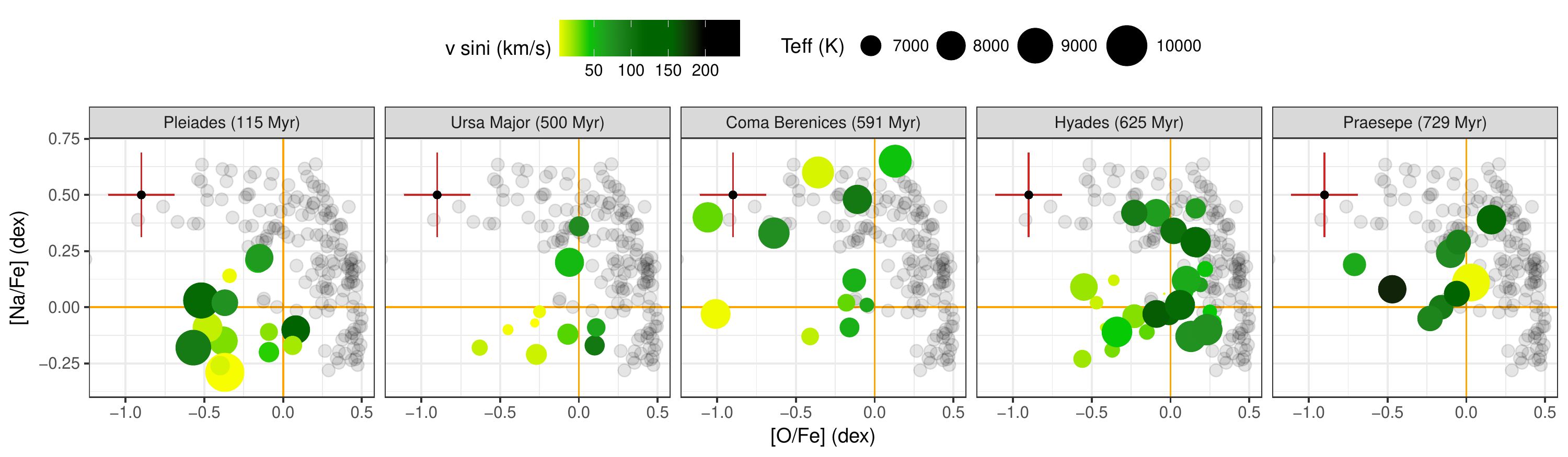}
   \includegraphics[width=\textwidth]{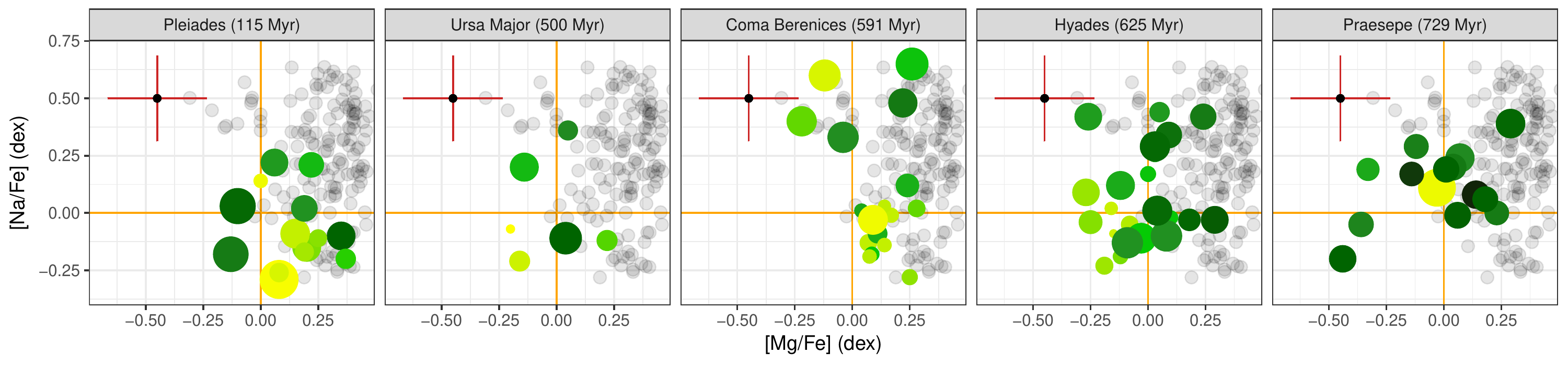}
   \caption{Top panels show the Na-O anti-correlation plane, bottom ones the
   Na-Mg one (Al measurements are not available). Reference data for GCs are
   plotted as grey points \citep{pancino17b}, the solar abundances as orange
   lines. Each OC star is coloured based on its projected rotational velocity
   and the size of the symbols reflects T$_{\rm{eff}}$ (hence, mass). Typical
   (median) errorbars are plotted as red crosses.} 
   \label{fig:nao}
    \end{figure*}

Figure~\ref{fig:nao} (top panels) shows the collected data in the Na-O
anti-correlation plane, sorted by OC age. The region occupied by GC stars in
this plane is represented using the Gaia-ESO data from \citet{pancino17b}. It is
important to keep in mind that GC anti-correlations tend to be less extended for
more metal-rich and less massive GCs \citep{carretta10,pancino17b}, and
therefore even if OCs had true anti-correlations, they would not be as extended
as in GCs. Additionally, and unlike in the GC case, all OCs have some stars in
the region of low Na and low O, typically close to the solar values\footnote{The
Am and Fm stars affected by strong metallicism, that were removed from the
sample (Section~\ref{sec:data} and Figure~\ref{fig:vsini}) occupy in the Na-O
plane the region where both [Na/Fe] and [O/Fe] are subsolar, mostly because they
have increased [Fe/H] abundances.}. This is unavoidable because the "normal"
population in OCs has Solar metallicity and no $\alpha$-enhancement, and this
could imply more vertical anti-correlations than in GCs, if they were present.

Having said that, the first thing that appears clear is that the distribution of
stars in the Na-O plane changes significantly from one OC to the other. The
Pleiades and Praesepe stars show no particular resemblance to the GC
anti-correlation patterns, apart from the bimodality in the Pleaides discussed
above.

In Coma Berenices, there is a clump of stars around solar abundance with
T$_{\rm{eff}} \leq 7000$~K (the "normal" stars) that is well separated from a
top distribution of hotter stars, with [Na/Fe] confined between 0.3 and 0.7~dex
and [O/Fe] spread between solar and --1.0~dex. The high-Na stars cover the same
area of the extreme populations observed for example in NGC~2808 \citep[see][and
references therein]{pancino17b}.

In the Hyades, the stars with $v \sin i \geq$50~km~s$^{-1}$ (dark green
symbols in Figure~\ref{fig:nao}) cover a large part of the less extreme Na-O
anti-correlation observed in typical GCs: the spread in [Na/Fe] is
$\simeq$0.6~dex, the spread in [O/Fe] $\simeq$0.5~dex, and the angular
coefficient of the anti-correlation is --0.47$\pm$0.11, overlapping the
reference GC data for the metal-intermediate to metal-rich GCs. In Ursa Minor
the stars rotating faster than $v \sin i \geq$50~km~s$^{-1}$ (dark green symbols
in Figure~\ref{fig:nao}) display a spread of $\simeq$0.5~dex in [Na/Fe] and of
$\simeq$0.3~dex in [O/Fe], with an anti-correlation angular coefficient of
--2.5$\pm$0.4, but the sample is smaller in this case.

To summarize, the fast rotating stars ($v \sin i \geq$50~km~s$^{-1}$) in the
Hyades and in Ursa Major display a Na-O anticorrelation, while the hottest stars
in Coma Berenices cover the extreme part of the GC anti-correlation, with high
Na and a range of O reaching extreme depletion, and a clump of "normal" stars
around solar values. These anti-correlations help understanding the
inversion of the peak heights (or of the skewness) in the [Na/Fe] and [O/Fe]
distributions. The [O/Fe] bimodality the Pleiades suggests that they also
might contain a Na-O anti-correlation, that might be revealed with larger
samples. Even considering all the due caveats on sample size and selection
function, these patterns do call for further investigation, in the light of a
possible connection with the MP problem in GCs.

\subsection{Magnesium depletion}

Magnesium is a key element for the MP phenomenon in GCs. It can only be
significantly destroyed in the Mg-Al hotter sub-cycles, that require about
80~10$^6$K to become efficient \citep[see figure~8 by][]{prantzos07}. At the
same time, it is not difficult to enhance Al in the Mg-Al cycle: because Mg is
so much more abundant, only a modest conversion of Mg in Al is sufficient to
significantly enhance [Al/Fe]. For this reason, the fact that the extent of the
Mg-Al anticorrelation is extremely variable with the GC mass and metallicity
\citep{pancino17b} is a very strong constraint on the possible production sites
of MPs in GCs, if one makes the hypothesis that the Mg-Al cycle is the sole
responsible for the observed patterns. Mg is also a problem: the very high
temperatures required are difficult to obtain in fast rotating massive stars
\citep{decressin07a,decressin07b}, and only a narrow range of asymptotic giant
branch star masses can produce the Mg-Al anticorrelation without destroying too
much Na \citep{ventura13,renzini15,prantzos17}. 

It is therefore very surprising to observe that the hottest stars
(Figure~\ref{fig:nao}, lower panels) in Coma Berenices cover the entire extent
of the Mg depletion observed in the most extreme GC stars (well below solar),
while at the same time having a high [Na/Fe] (up to 0.7~dex). In fact, A stars
in Coma Berenices have a typical mass of $\sim$2~M$_{\odot}$ and thus cannot
activate efficiently the Mg-Al cycle in their cores. In the
Hyades and in Praesepe, the Na-rich stars also cover a range in Mg, with
similarly extreme depletion, but in that case there are also several stars with
low Na and a range of Mg, that might be chemically peculiar or point to a
different chemistry altogether.

It is difficult to measure Al in these stars (I use Na in Figure~\ref{fig:nao}),
and we do not expect significant Al variations in metal-rich GCs
\citep{pancino17b}, but \citet{fossati11} did measure Al for 6 stars in NGC~5460,
a poorly studied OC of $\simeq$160~Myr. The result is an Al variation of about
0.3~dex, accompanyed by an Mg variation largely exceeding 1~dex. This, among
other facts, supports the idea that the low Mg displayed by all OCs examined here
is not caused by the Mg-Al cycle burning, but by diffusion and rotational mixing
as described in Section~\ref{sec:data}. It is therefore extremely interesting,
and worthy of further investigation, that the observed chemistry resembles (at
least in part) what is observed in the extreme population of GCs. This piece of
evidence becomes even more suggestive when considered together with the Na
bimodality and the Na-O anti-correlations discussed above.

\subsection{Other elements}
\label{sec:other}

It would be extremely interesting to study the C-N anti-correlation plane as
well, but unfortunately N is not provided in the examined works, except for a
handful of stars in Praesepe. The behaviour of carbon with T$_{\rm{eff}}$ and $v
\sin i$ is indeed qualitatively similar to that of oxygen, with a similar or
more extreme depletion, depending on the OC. 

Helium is a key element for MPs in GCs. If the MP chemistry is produced in CNO
burning and related hotter cycles, we expect the He abundance to vary as well,
with peculiar stars having higher He \citep[see][and references
therein]{bastian15}. Helium measurements are not common in the stars analyzed
here. The theoretical expectations -- with all other parameters fixed -- predict
an increase of the He surface abundance with increasing rotational velocity
\citep{ekstrom12}. A dedicated He study, for example in correlation with Na and
Mg, would be extremely interesting.

Lithium is also observed to vary among GC stars, anti-correlating with Na or Al
\citep[see][and references therein]{dorazi15}. This is a problem because the
proposed polluters do not reproduce the observed chemistry well. All the stars
analyzed here are hotter than the Li dip, a dramatic drop in Li abundance
occurring around 6700~K \citep{boesgaard16}, with only the coolest F stars
nearing it. Theoretical expectations are that Li gets progressively depleted
with increasing rotation \citep{pinsonneault89}, and the Li dip can be explained
as the final outcome of Li destruction mediated by internal mixing and diffusion
mechanisms. Unfortunately, the behaviour of Li with rotation for A-type stars is
not so well known: the Li line is not observed in fast rotators much hotter than
the dip, because it is weak and further weakened by rotation. 

Finally, s-process elements (Yr, Zr, Ba) in the studied stars are enhanced up to
0.5~dex and in some cases to 1~dex, and this is especially true for Am stars,
displaying in general the highest s-process element abundances. This is also in
line with what observed in some GCs.

There are also some elements which do not vary in the vast majority GCs,
like heavier $\alpha$-elements (Ca, Ti), or most iron-peak elements (Fe, V, Sc,
Ni, Cr) that indeed vary in Am and Fm stars, and can reach extreme variations,
like those shown in Figure~\ref{fig:vsini} (metallicism). One clear indicator of
metallicism, besides an enhancement in iron-peak elements is a very low scandium
abundance \citep{alecian13}. These stars are unlikely to have implications for
the MP problem, are never observed in GCs,  and were not studied here. They are
generally slow rotators, in which rotation cannot inhibit the diffusive
mechanisms.

In summary, the effect of rotation -- or better, of the interplay between
rotational mixing and diffusion processes -- appears to change all the
GC-relevant elements on the surface of these relatively cool stars, and
generally in the right direction, even if they are not altered by CNO nuclear
processing in these stars at all.


\section{Implications for globular clusters}
\label{sec:gc}

The main results discussed so far can be summarized as follows:
\begin{enumerate}\item{four of the five OCs display an apparent bimodality
in [Na/Fe], with the low-Na peak more populous than the high-Na peak
(Figure~\ref{fig:hist}); three out of five OCs pass a BIC test for
bimodality;}\item{A possible [O/Fe] bimodality is also apparent, although
noisier (only two OCs pass the BIC test), with an inversion of peak
population (or of skewness) compared to the [Na/Fe] distribution;}\item{Two
OCs, Ursa Major and the Hyades, display a Na-O anti-correlation among stars with
$v \sin i \geq$50~km~s$^{-1}$ (Figure~\ref{fig:nao}), that explains the
inversion of peak population mentioned before; this suggests that also the
Pleiades might contain a Na-O anti-correlation, given that they show an
[O/Fe] bimodality;}\item{The last OC, Coma Berenices, contains in the high-Na
peak stars with a range of O and Mg abundances, that reache the extreme
depletion of NGC~2808, and the low Mg is clearly not produced by Mg-Al cycling.}
\end{enumerate}

While these results need of course to be double-checked and further
investigated, this is the first detection of a Na-O anti-correlation in OC dwarf
stars. Unfortunately, two complications prevent a direct and obvious connection
with the GC case. The first is that while rotation appears to be an important
ingredient of the observed Na-O anti-correlation and bimodality, because only
stars rotating faster than $v \sin i \geq$50~km~s$^{-1}$ have clear Na-O
anticorrelations, the exact ingredient that separates stars in Na-rich and
Na-poor is not identified yet (Section~\ref{sec:hist}). The second is that the
observed chemical patterns are too superficial in these stars
\citep[10$^{-6}$--10$^{-4}$ of the stellar mass,][]{richard02} and are not
expected to survive the first-dredge up when they will start ascending the red
giant branch, unlike GC stars. This is also confirmed observationally, because
generally OC giants do not display Na-O anti-correlations
\citep{desilva09,smiljanic09,pancino10,carrera11,carrera13,maclean15}. Therefore
this new piece of evidence does not provide immediate answers, for the moment it
just poses more questions, and an indication that we should explore more deeply
the role of stellar rotation as a promising avenue to interpret the MP
phenomenon in GCs (and OCs).

That rotation is important for stellar and cluster evolution, and that it
might even play a role in the MP phenomenon, is obvious by the vast body of
literature already available on the subject. We do know that the extended
turn-off observed in several Magellanic Cloud clusters with age below 2~Gyrs
could be explained with differential stellar rotation
\citep{bastian09,niederhofer15}, while above 2~Gyrs MPs appear on the red giant
branch \citep{martocchia18}. We also know that the Hyades and Praesepe do
display an extended turn-off as well \citep{brandt15}. Rotation also is known to
alter stellar structural properties like radius and T$_{\rm{eff}}$
\citep{somers17} and thus the subsequent stellar evolution. Stellar rotation
disappears along the red giant branch, but internal rotation is observed among
red giants \citep[with asteroseismology,][]{corsaro17}, and it is expected to
partially resurface in the helium burning phase, when the star has lost some
more mass and is more compact. Indeed, differential rotation is observed along
the blue horizontal branches of GCs \citep{behr03,recio04} and it varies with
the position along the branch, similarly to the He, Na, and s-process abundances
\citep{marino11,marino13,marino14}. One of the most explored scenarios for MPs
indeed relies on fast rotating massive stars \citep{decressin07b} as polluters.

But as mentioned above, rotation alone does not explain the observed
bimodalities (Figure~\ref{fig:hist}). Among the various other physical phenomena
that merit deeper investigation in the framework of MPs, there is also stellar
binarity and especially close binary interactions, which could act in various
ways. Firstly, binary stars are very numerous along the main sequence
\citep{demarco17}, ranging from at least 80\% in O and B stars, to at least
50--60\% in F and A stars, respectively, with no detected difference between
(open) cluster and field environments. Secondly, binarity can alter the rotation
and mass loss properties of stars, possibly enhancing rotational mixing effects
\citep{chatzopoulos12}. Thirdly, binary interactions can lead to a variety of
exotic results in the dense cluster environments, like blue stragglers
\citep{sandage53}, red stragglers and sub-subgiants, cataclismic variables,
X-ray binaries \citep{cool13,geller17}. In cluster environments, binaries are
formed, destroyed, and stellar mergers and mass transfer episodes can occur
\citep{benacquista13}. Indeed, the binary fraction seems to be lower in GCs and
decreases with increasing GC mass \citep{milone17}, which is also one of the
driving parameters of the anti-correlation extension
\citep{carretta10,pancino17b}. The binary fraction of enriched stars in GCs
appears lower \citep{lucatello15}, pointing towards a possible higher binary
destruction or merger rate in enriched stars. All these phenomena add
stochasticity, that could produce the differences observed from GC to GC, and
potentially help in the dense GC environment to produce exotic chemistry,
possibly able to persist along stellar evolution, as compared to OCs.

To conclude, rotation and binarity could be the two missing ingredients
that -- combined together with relatively cool CNO burning and other diffusion
processes -- could bring a solution to the long-standing MP problem in GCs.
These are complex processes to model in a self-consistent way, but the new
astroseismology results from the CoRoT \citep{michel08} and Kepler
\citep{borucki10} missions are stimulating theoretical work in this sense and
therefore we might be closer to having all the needed tools to finally solve the
MP problem in GCs.


\begin{acknowledgements} EP would like to warmly thank the following colleagues,
for enlightening discussions about various aspects touched in this paper:
G.~Altavilla, N.~Bastian, I.~Cabrera-Ziri, C.~Charbonnel, M.~Fabrizio,
E.~Franciosini, M.~Gieles, V.~Henault-Brunet, R.~Izzard, C.~Lardo, L.~Magrini,
S.~Marinoni, C.~Mateu, A.~Mucciarelli, S.~Randich, V.~Roccatagliata, G.~Sacco,
M.~Salaris, N.~Sanna, and A.~Sills. EP would also like to thank an
anonymous referee, who helped in making the results presentation and the
discussion clearer. This research has made extensive use of the NASA ADS
abstract service, the arXiv astro-ph preprint service, the CDS Simbad and Vizier
resources, the Topcat catalogue plotting tool \citep{taylor14}, and the R
programming language and R Studio environment. \end{acknowledgements}


\bibliographystyle{aa}
\bibliography{HyadesNaO}


\end{document}